\documentclass[preprint2]{aastex}
\usepackage{graphicx}

\shorttitle{NIR Spectra of Cas~A \& Kepler}
\shortauthors{Gerardy \& Fesen}
\slugcomment{Submitted to The Astronomical Journal}

\begin{document}
\title{Near-Infrared Spectroscopy of the Cassiopeia~A \\ and Kepler Supernova 
Remnants}
\author{Christopher L. Gerardy and Robert A. Fesen}
\affil{6127 Wilder Laboratory, Department of Physics \& Astronomy, \\ 
	Dartmouth College, Hanover, NH 03755-3528}

\begin{abstract}
Near-infrared spectra (0.95 -- 2.4 \micron) of the Cassiopeia~A and Kepler
supernova remnants (SNRs) are presented. Low-dispersion (R $\approx$ 700) 
spectra were obtained for five bright fast-moving ejecta knots (FMKs) at two 
locations on the main shell and for three bright circumstellar knots (QSFs) near
the southwest rim of Cas~A.  The main shell FMKs in Cas~A exhibit a 
sparse near-infrared spectrum dominated by [\ion{S}{2}] 1.03 \micron\ 
emission with a handful of other, fainter emission lines.  Among these
are two high-ionization silicon lines, [\ion{Si}{6}] 1.96~\micron\ and 
[\ion{Si}{10}] 1.43~\micron, which have been detected in AGNs and novae but 
never before in a supernova remnant.  The near-infrared spectra of circumstellar
QSFs in Cas~A show a much richer spectrum, with strong \ion{He}{1} 
1.083~\micron\ emission and over a dozen bright [\ion{Fe}{2}] lines.  Observed 
[\ion{Fe}{2}] line ratios indicate electron densities of 
5 -- 9 $\times$ 10$^{4}$~cm$^{-3}$ in the QSFs.  The Cas~A QSF data are quite 
similar to the observed spectrum of a bright circumstellar knot along the 
northwest rim of the Kepler SNR, which also 
shows strong \ion{He}{1} and [\ion{Fe}{2}] emission with a measured electron 
density of 2.5 -- 3 $\times$ 10$^{4}$~cm$^{-3}$.  Finally, we present 
\textit{J}- and \textit{K}-band images of Cas~A.  The \textit{K}-band image 
shows faint diffuse emission which has no optical or mid-infrared counterpart 
but is morphologically similar to radio continuum maps and may be infrared 
synchrotron radiation.
\end{abstract}

\keywords{supernova remnants --- ISM: individual (Cassiopeia~A, Kepler SNR) --- 
ISM: lines and bands --- circumstellar matter --- infrared radiation}

\section{Introduction}
The numerous optical and UV spectroscopic studies of supernova remnants (SNRs) 
in the literature have resulted in a rich catalog of observed emission-line 
features (cf. Fesen \& Hurford 1996 and references therein).  However, most of 
these studies do not extend much beyond 8500 \AA\ due to the poor 
sensitivity of optical detectors farther to the red.   Only a handful of 
published SNR spectra go out as far as 1.1 \micron\ where the most commonly 
seen features are [\ion{S}{3}] 9069, 9532~\AA, [\ion{C}{1}] 9823, 9850~\AA, 
[\ion{S}{2}] 10287--10372~\AA, and \ion{He}{1} 10830~\AA\
\citep[e.g.,][]{DA81,D82,HMK84}.  

With the maturing of high quantum efficiency near-infrared (NIR) 
detectors, it is now possible to probe the spectra of SNRs in the 
1 -- 5~\micron\ wavelength regime, and there are a number of reasons to do so.  
Near-infrared spectroscopy can provide access to features that aren't 
available in optical and UV spectra, such as molecular species like H$_{2}$ and 
CO, which can provide valuable information about molecule formation and 
destruction in SNRs.  Molecular emission can also often be used as an 
astrophysical probe, providing information about temperature, density, and 
excitation mechanisms.  In addition, there are many strong lines of 
[\ion{Fe}{2}] in the NIR, which can be used as density diagnostics.  These lines
have higher critical densities than optical density sensitive lines, and thus 
can probe much higher densities ($\sim 10^{3}$ -- $10^{5}$ cm$^{-3}$).  Finally,
the near-infrared contains several strong high-ionization species which have 
been seen in other objects like AGNs, novae, and planetary nebulae.  These lines
could provide important information about ionization processes and structures in
supernova remnants.  

Unfortunately, few NIR spectra of supernova 
remnants currently exist in the literature. Furthermore, there are almost no 
near-infrared data that take advantage of large-format array detectors allowing 
for spatially resolved long-slit spectroscopy and broad wavelength coverage.  
Indeed, of the few datasets available, most were obtained with relatively large 
apertures ($\sim$5\arcsec -- 20\arcsec) in narrow wavelength regions typically
covering only a single emission feature with each bandpass.  

To date, the most comprehensive work on the NIR spectra of supernova remnants 
has been the study of molecular hydrogen (H$_{2}$) emission from shocks running 
into molecular clouds. This has been seen in IC~443, RCW~103, and the Cygnus 
Loop \citep{T79,GWL87,OMD89,OMD90,BS93,GWHL91,GWG91,RGW95}.  H$_{2}$ has also 
been detected in the Crab Nebula \citep{GWL90}, presumably from dense, neutral 
cores of the emission-line filaments.

Other near-infrared work on SNRs has focused on [\ion{Fe}{2}] line emission, 
which is typically two orders of magnitude brighter, relative to \ion{H}{1}, 
than that seen in \ion{H}{2} regions \citep[hereafter OMD89]{SHMC83,OMD89}.  
The large [\ion{Fe}{2}]/\ion{H}{1} ratios observed have been suggested as a 
good tracer for shocks in extragalactic studies (e.g., OMD89), and near-infrared
[\ion{Fe}{2}] imaging has been used to probe extragalactic SNR populations 
\citep[e.g.][]{G97}. However, a large [\ion{Fe}{2}]/\ion{H}{1} ratio is not 
always an indication of shocks as the strong [\ion{Fe}{2}] emission seen 
in the Crab Nebula is likely a result of photoionization  \citep{GWL90}.  
Also, relative NIR [\ion{Fe}{2}] line ratios have been used to deduce 
electron densities for Kepler, N63A, N49, N103B (OMD89), RCW~103 
\citep[OMD89;][hereafter OMD90]{OMD90}, and the Crab Nebula \citep{RRP94}. 

In this paper, we present 0.95 -- 2.4~\micron\ spectra of shocked,
metal-rich ejecta in the Cassiopeia~A (Cas~A) supernova remnant as well as 
\textit{J}- and \textit{K}-band images.  To our knowledge, these are the first 
published NIR spectra of a young, ``oxygen-rich'' supernova remnant.  We also 
present NIR spectra of shocked circumstellar mass-loss material, both in Cas~A 
and in Kepler's SNR. 

\section{Observations and Data Reduction}

Low-dispersion near-infrared spectroscopy and imaging of 
the Cas~A and Kepler supernova remnants were obtained with the 2.4m 
Hiltner telescope at MDM Observatory on the southwest ridge of Kitt Peak in 
Arizona.  Spectroscopic observations of Cas~A took place in late November and 
early December 1999. \textit{J}- and \textit{K}-band images of Cas~A were 
obtained in November 2000.  Kepler's SNR was observed in April 2000.   

All observations were obtained with TIFKAM (a.k.a.\ ONIS), a high-throughput 
infrared imager and spectrograph with an ALLADIN 512 $\times$ 1024 InSb 
detector.  This instrument can be operated with standard \textit{J}, \textit{H},
and \textit{K} filters for broadband imaging, or with a variety of grisms, 
blocking filters, and an east-west oriented $0\farcs 6$ slit, allowing low 
($R \approx 700$) and moderate ($R \approx 1400$) resolution spectroscopic 
observations from 0.95 to 2.5~\micron. 

\textit{J}- and \textit{K}-band imaging of Cas~A was performed using the 
following procedure:  Sets of four dithered 30~s on-target images were 
immediately followed by four dithered 30~s images of fields $\approx$ 
10\arcmin\ off-target. The off-target images were averaged together with 
high-pixel rejection to remove stars, creating sky background images which were 
then subtracted from the on-target images.  This process was repeated four times
for each on-target pointing. The entire remnant was covered in several 
overlapping positions resulting in total on-target integration times of 32 -- 64
minutes in \textit{K}-band, and 24 -- 48 minutes in \textit{J}-band.  
Sky-subtracted on-target images were registered and combined using standard 
IRAF tasks.  \textit{J}-band imaging of the Kepler SNR was performed in a 
similar manner, but with only one on-target position and 8 minutes total 
on-target integration time.  

\begin{figure}[t]
\begin{center}
\includegraphics[scale=0.375,keepaspectratio=true,angle=0]{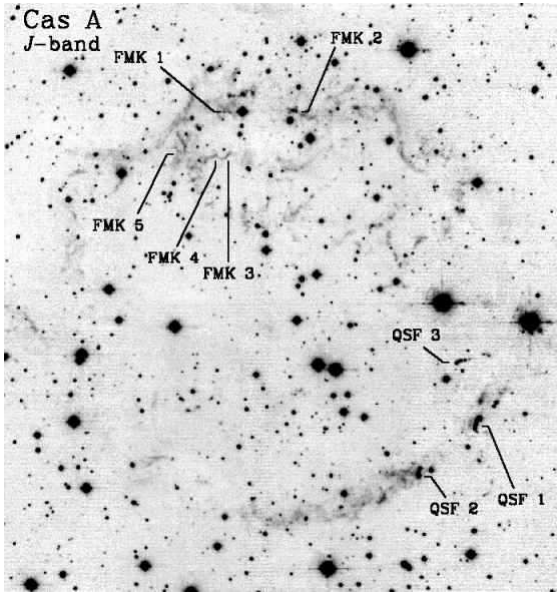}
\caption{\protect\textit{J}-band image of Cas~A with the locations
of the observed FMKs and QSFs marked.  North is up and East is to the left.
\label{casajim_lab}}
\end{center}
\end{figure}

\begin{figure}[t]
\begin{center}
\includegraphics[scale=0.375,keepaspectratio=true,angle=0]{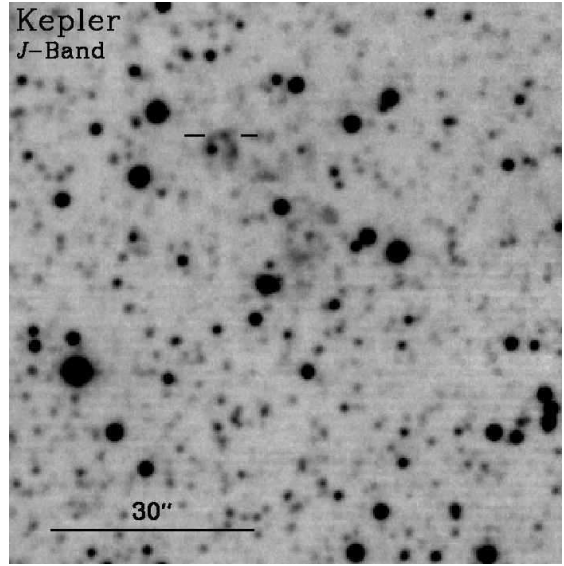}
\caption{\protect\textit{J}-band image of Kepler with the location
of the observed circumstellar knot marked.  North is up and East is to the left.
\label{kepjim}}
\end{center}
\end{figure}

Near-infrared 0.95 -- 2.4~\micron\ long-slit spectra of Cas~A and Kepler were 
obtained using three spectroscopic setups covering the 0.95 -- 1.8~\micron, 
1.2 -- 2.2~\micron, and 2.0 -- 2.4~\micron\ wavelength regions.  In Cas~A, five 
metal-rich ejecta knots (FMKs: ``Fast-Moving Knots'') were observed at two slit 
positions on the main shell and three bright knots of circumstellar mass-loss 
material (QSFs: ``Quasi-Stationary Flocculi'') were observed near the southwest 
rim of the remnant.  The Kepler SNR was observed at a single position on the 
northwest rim, the region with the brightest optical emission.  The 
spectroscopically observed regions of Cas~A and Kepler are marked on the
\textit{J}-band images shown in Figures \ref{casajim_lab} and \ref{kepjim}.

\begin{figure*}[t]
\begin{center}
\includegraphics[scale=0.55,keepaspectratio=true,angle=-90]{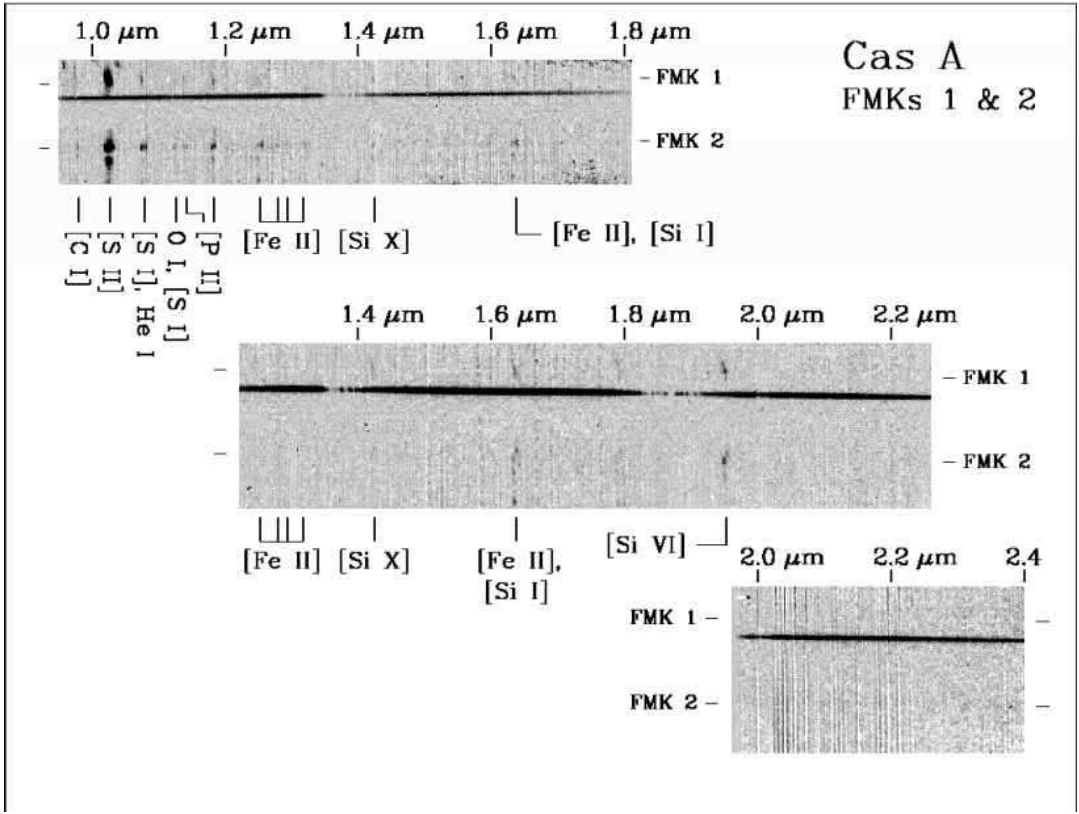}
\caption{2-D long-slit spectra of Cas~A FMKs 1 \& 2.
 Each frame shows a single 900~s exposure from a spectroscopic
setup.  Full 0.95--2.4 \protect\micron\ coverage was achieved with three overlapping
spectroscopic setups.  First order removal of night-sky lines has been
performed, but the data shown are not corrected for telluric absorption or
instrumental response.  The approximate wavelength scale (in \protect\micron) is
marked along the top of each frame, and the positions of the various
observed features are shown at the bottom. Lines at the left and right of each
frame denote the ends of the 1-D extraction apertures.\label{2dfmk}}
\end{center}
\end{figure*}

\begin{figure*}[t]
\begin{center}
\includegraphics[scale=0.55,keepaspectratio=true,angle=-90]{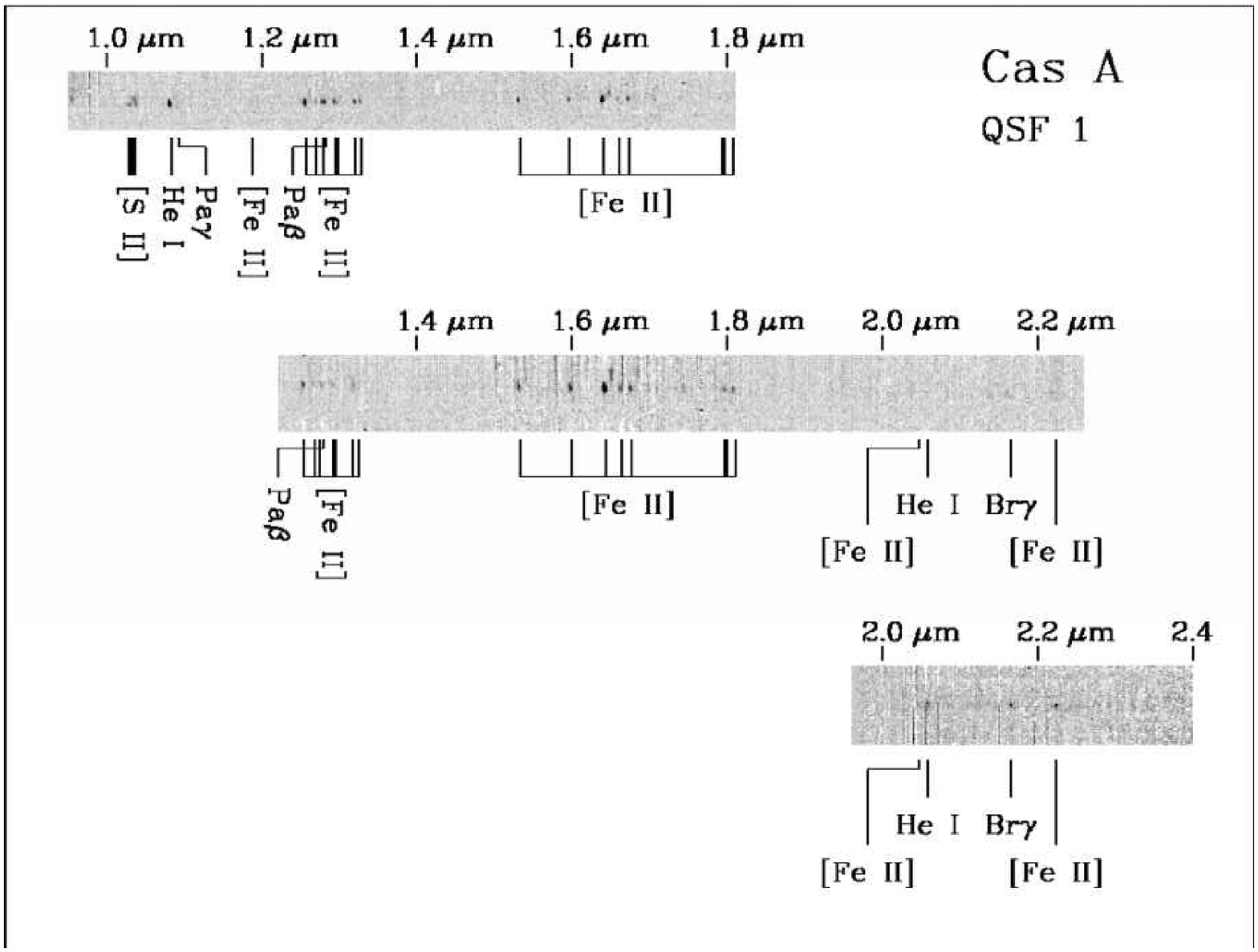}
\caption{2-D long-slit spectra of QSF 1 in Cas~A.
Each frame shows a single 900~s exposure from a spectroscopic
setup.  Full 0.95--2.4 \protect\micron\ coverage was achieved with three overlapping
spectroscopic setups.  First order removal of night-sky lines has been
performed, but the data shown are not corrected for telluric absorption or
instrumental response.  The approximate wavelength scale (in \protect\micron) is marked
along the top of each frame, and the positions of the various observed
features are shown at the bottom. [\protect\ion{S}{2}] and [\protect\ion{Fe}{2}] emission from a
faint FMK can be seen just above the QSF spectrum.\label{2dqsf}}
\end{center}
\end{figure*}

Table \ref{speclog} lists the log of spectroscopic exposures for each slit 
position. Between on-target exposures the telescope was dithered along the slit 
to sample the array at multiple locations.  For sufficiently isolated knots 
(QSF 1, QSF 3 and Kepler), dithered on-target exposures were used for 
first-order night-sky subtraction.  For the other regions (QSF~2 and the FMKs), 
night-sky spectra were obtained between on-target exposures by observing an 
empty location 10\arcmin\ north of the remnant and then removed from the 
on-target data.  Figures \ref{2dfmk} and \ref{2dqsf} show representative 2-D 
long-slit spectra 
of FMKs 1 \& 2 and QSF~1 in Cas~A.  Each frame shown is a single 900~s exposure 
in a single spectroscopic setup after first-order night-sky subtraction has been
performed.  1-D spectra were extracted from these 2-D frames using standard IRAF
tasks.  Arc-lamps were observed at each telescope position to provide 
wavelength calibration.  

The spectra were corrected for telluric absorption by
observing nearby A~stars and early G~dwarfs from the Bright Star
Catalog \citep{BSC}.  Applying the procedure described by
\citet[hereafter HRL98]{HRL98}, stellar features were removed from
the G~dwarf spectra by dividing by a normalized solar spectrum
\citep{solar1, solar2}\footnote{NSO/Kitt Peak FTS data used
here were produced by NSF/NOAO.}. The resulting spectra were
used to correct for telluric absorption in the A~stars.  The hydrogen
features in the corrected A~star spectra were removed from the
raw A~star spectra and the results were used to correct the target
data for telluric absorption. [For further discussion of this procedure
see HRL98; \citet{HCR96}, and references therein.] The instrumental
response was calibrated by matching the continuum of the A~star telluric
standards to the stellar atmosphere models of \citet{kurucz}.

After correction for instrumental response and telluric absorption, 
the data for each knot in given spectroscopic setup were averaged
together.  Data taken with different setups were then flux-matched in the 
overlapping wavelength regions, and joined together to make a single, full 
coverage 1-D spectrum.  Three \citet{MassGron} spectrophotometric standards were
observed to set the absolute flux levels.  The resulting absolute flux 
calibration is believed accurate to $\simeq$ 20\% shortward of 1.8~\micron, and 
$\simeq$ 30\% from 1.8 to 2.4~\micron.

\section{Results \& Discussion}
\begin{figure}[t]
\begin{center}
\includegraphics[scale=0.30,keepaspectratio=true,angle=-90]{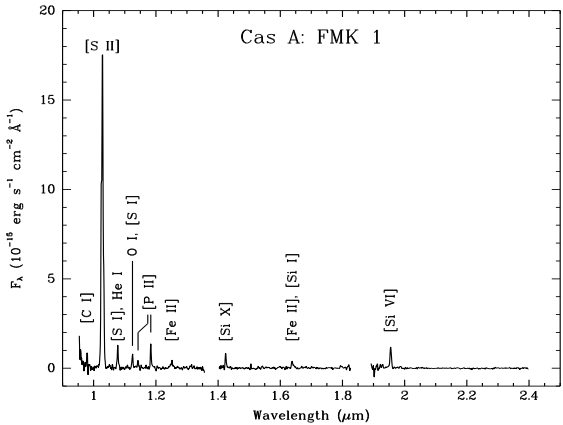}
\caption{Observed NIR spectrum of FMK 1 in Cas~A. The
spectrum is dominated by strong [\protect\ion{S}{2}] emission, but also exhibits a
number of other faint lines including [\protect\ion{C}{1}], [\protect\ion{P}{2}], [\protect\ion{Fe}{2}],
and high-ionization lines of [\protect\ion{Si}{6}] and [\protect\ion{Si}{10}]. \label{FMK1}}
\end{center}
\end{figure}

\begin{figure}[t]
\begin{center}
\includegraphics[scale=0.30,keepaspectratio=true,angle=-90]{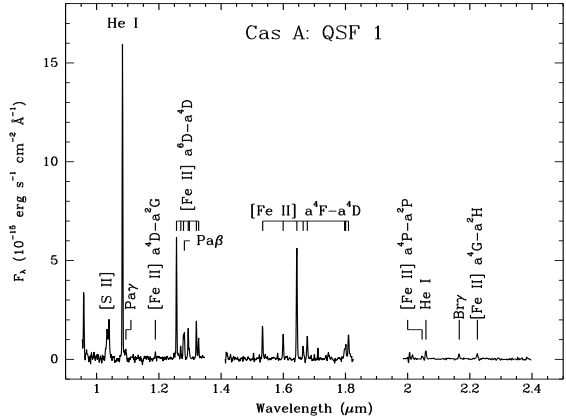}
\caption{Observed NIR spectrum of QSF 1 in Cas~A.  The
spectrum is dominated by strong \protect\ion{He}{1} 1.083 \protect\micron\ and [\protect\ion{Fe}{2}]
emission lines. \label{QSF1}}
\end{center}
\end{figure}

\begin{figure}[t]
\begin{center}
\includegraphics[scale=0.30,keepaspectratio=true,angle=-90]{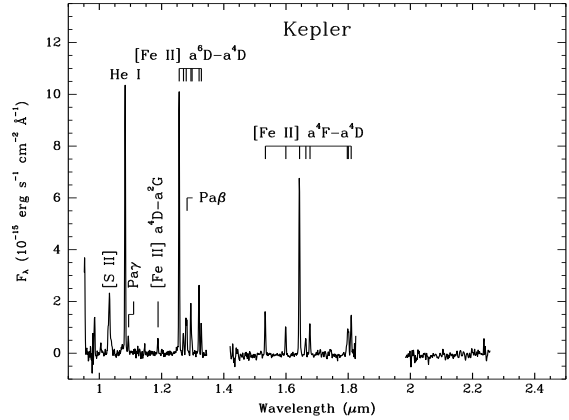}
\caption{Observed NIR spectrum of a bright circumstellar
knot on the west rim of Kepler's SNR.  The spectrum is dominated by strong
\protect\ion{He}{1} 1.083 \protect\micron\ and [\protect\ion{Fe}{2}] emission lines. \label{kepler}}
\end{center}
\end{figure}

Observed spectra of FMK 1 and QSF 1 in Cas~A are presented in Figures \ref{FMK1}
and \ref{QSF1} respectively with the spectrum of the west rim of the Kepler SNR
shown in Figure \ref{kepler}.  Line identifications for the 
Cas~A FMK spectra are presented in Table~\ref{FMK12tab} (FMKs 1 \& 2) and 
Table~\ref{FMK345tab} (FMKs 3, 4 \& 5) along with measured line centers 
and line fluxes, both observed and dereddened.  Table~\ref{QSFkeptab} shows the
line identifications and observed and dereddened fluxes for the Cas~A QSFs
and the western knot in Kepler.  All wavelengths are given as vacuum values.  
Dereddening of the observed spectra was performed using the extinction curve 
of \citet{CCM89}.  For the Cas~A data, \textit{E(B--V)} = 1.5 was used although 
the actual extinction varies significantly across the remnant 
\citep[hereafter HF96]{HF96}.  For the Kepler data, we assumed 
\textit{E(B--V)} = 0.9 \citep{BLV91} 

\subsection{Cas~A FMK Spectra}

Optical spectra of Cas~A ejecta knots (FMKs) exhibit strong forbidden 
oxygen and sulfur emission, and contain a number or fainter 
metal lines.  These spectra show no indication of H or He, and little, if 
any, [\ion{Fe}{2}] emission (HF96). (Note: Complete 4000 -- 10500~\AA\ optical
spectra of FMKs 1 and 2 are presented by HF96.) The near-infrared spectra of 
Cas~A FMKs are 
intrinsically faint compared to their optical emission, and they are dominated 
by the [\ion{S}{2}] 1.03~\micron\ blend. A handful of other faint, 
low-ionization forbidden lines are seen including [\ion{C}{1}] 0.9827~\micron, 
0.9853~\micron, [\ion{P}{2}] 1.1471~\micron, 1.1886 \micron, and [\ion{Fe}{2}] 
1.2570~\micron, 1.6440~\micron.  We note that the 1.6440~\micron\ feature could 
be blended with [\ion{Si}{1}] 1.646~\micron\ emission. In FMK~2, the brightest 
FMK we observed in Cas~A, three other faint [\ion{Fe}{2}] lines 
(1.2791~\micron, 1.2946~\micron, \& 1.3209~\micron) were weakly detected.  Two 
other weak lines near 1.08~\micron\ and 1.13~\micron\ could be emission from 
the [\ion{S}{1}] \mbox{$^{3}$P--$^{1}$D} doublet, but could also be due to (or 
blended with) \ion{He}{1} 1.083~\micron\ and \ion{O}{1} 1.129~\micron.  
\ion{O}{1} emission would be consistent with the presence of weak permitted 
\ion{O}{1} lines seen in the optical spectra of Cas~A FMKs (e.g. HF96). 

Perhaps the most interesting near-infrared lines detected in the brighter FMKs 
(1, 2, \& 4) are those near the rest wavelengths of 1.43~\micron\ and 
1.96~\micron.  We identify these as high-ionization lines of silicon, 
[\ion{Si}{6}] 1.965~\micron\ and [\ion{Si}{10}] 1.4305~\micron.  The detection 
of such high ionization species was unexpected as no other high-ionization lines
are observed in optical or near-infrared spectra of Cas~A. However, we could
find no other likely line identifications that fit the observed wavelengths.  
There are weak H$_{2}$ transitions at wavelengths near both lines, but we find 
no evidence of emission present from the much stronger H$_{2}$ feature at 
2.12~\micron.  Also, the observed Doppler shifts of these lines, if identified 
as [\ion{Si}{6}] and [\ion{Si}{10}], are consistent with each other and match 
those of the other lines seen in the FMKs to within the resolution of our data 
($\approx$ 450 km~s$^{-1}$).  The absence of these lines in the spectra of the 
fainter FMKs is not significant as both of these lines suffer moderate telluric 
absorption and they could be too weak to detect in these fainter knots. 

This is the first detection of these high-ionization silicon features in the 
spectrum of a supernova remnant.  OMD90 looked for, but did not detect 
[\ion{Si}{6}] in their spectra of RCW~103. However, the [\ion{Si}{6}] 
1.965~\micron\ line has been seen in planetary nebulae \citep{AH88}, and both 
the [\ion{Si}{6}] 1.965~\micron\ and [\ion{Si}{10}] 1.4305~\micron\ lines have 
been reported in NIR spectra of the solar corona \citep{MNM67}, novae 
\citep{BD90,G90}, and AGNs \citep{OM90,T96,M00}.  Another
high-ionization silicon line, [\ion{Si}{7}] 2.48~\micron, has also been observed
in many of these objects, but this line was outside the observed bandpass of
the Cas~A spectra shown here.  However, in follow-up observations made in 
November 2000 this line was detected in 1.4 -- 2.5~\micron\ spectra of 
FMKs~1~\&~2.

NIR spectra of novae often show other strong high-ionization features 
such as [\ion{S}{9}] 1.252~\micron, [\ion{S}{9}] 1.392~\micron, 
[\ion{Cr}{11}] 1.550~\micron, [\ion{P}{8}] 1.736~\micron, [\ion{Al}{9}] 
2.040~\micron, and [\ion{Ca}{8}] 2.323~\micron\ \citep{WD96}, none of which are 
seen in our Cas~A data.  (Note: [\ion{S}{9}] 1.252~\micron\ 
would be blended with [\ion{Fe}{2}] 1.257~\micron, and [\ion{S}{9}] 
1.392~\micron\ is obscured by strong telluric absorption).  On the other hand, 
in near-infrared AGN spectra the [\ion{Si}{6}], [\ion{Si}{7}], and 
[\ion{Si}{10}] lines are often the only strong high-ionization features seen in 
the 0.95 -- 2.5~\micron\ region.  

Near-infrared ``coronal'' line emission in novae, AGNs, and
planetary nebulae is believed to be due to photoionization, although
collisional ionization from hot, shocked gas often cannot be ruled out.
In contrast, the low-ionization optical spectra of Cas~A's FMK knots 
have been modelled as shocks with a photoionized precursor component 
(Sutherland \& Dopita 1995; HF96).  Observed ratios of lines at different 
ionization levels are explained in these models as a mixture of emission from
the rapidly cooling post-shock region, and from a precursor in the ejecta out
ahead of the reverse-shock which is photoionized by UV emission from the shock
front.  Unfortunately, no predictions for [\ion{Si}{6}] and [\ion{Si}{10}] 
emission
have been made so it is unclear if the shock/precursor models can explain these
very high ionization lines.  Alternatively, the high-ionization emission might 
be understood with a pure photoionization model like those used to explain the 
high-ionization lines seen in novae, AGNs and planetary nebulae.  The main shell
of Cas~A is quite bright in X-rays \citep{H00,HHP00} which could provide a 
strong photoionizing source.

In any case, it also seems likely that abundances play some role in the 
detection of these high-ionization silicon lines.  Spectral modeling of optical 
FMK spectra indicate significant enrichment of metal abundances, suggesting that
FMKs are gas from the inner regions of the exploded star 
\citep[HF96]{ck79}.  Furthermore, X-ray imaging of Cas~A has detected 
bright silicon line emission from the main shell ejecta \citep{HHP00}.  All of 
this indicates that the FMKs are likely silicon-rich and so it may be that 
only a small fraction of the silicon is highly ionized with the bulk of the gas 
in a much lower ionization state.  However, without a detailed spectral model
it is impossible to conclude that the detection of these silicon lines is 
merely an abundance effect as these lines are also detected in AGNs and 
planetary nebulae where the silicon abundance is much lower.  In fact, with 
respect to the observed high-ionization lines, the Cas~A spectra more closely 
resemble near-infrared AGN spectra than NIR spectra of noave, even though the 
gas in novae is probably closer in composition to the FMKs. 

\subsection{Cas~A QSFs and Kepler}

The optical spectra of shocked circumstellar knots in Cas~A (QSFs) and Kepler's
SNR are dominated by strong [\ion{N}{2}] emission, with only a few other weak
lines of H and He.  These knots are thought to be nitrogen enriched material
shed by the progenitor star prior to the explosion, and then shocked by the 
expanding blast wave \citep{PV71}.  The NIR spectra of these knots seem 
consistent with this picture, exhibiting an ISM-like spectrum with the strong 
[\ion{Fe}{2}] emission typical of shocked gas.  The Cas~A QSFs and the Kepler 
knot are relatively bright in the near-infrared, especially in the \textit{J} 
($\sim$1.2~\micron) and \textit{H} ($\sim$1.6~\micron) bands due to this rich 
[\ion{Fe}{2}] spectrum.  In addition to the [\ion{Fe}{2}] lines, these 
circumstellar knots also show 
strong \ion{He}{1} 1.083~\micron\ emission, as well as the [\ion{S}{2}] 
1.03~\micron\ blend, [\ion{C}{1}] 0.9827~\micron, 0.9853~\micron, 
\ion{He}{1} 2.058~\micron, and faint hydrogen lines (Pa$\gamma$ 1.0941~\micron, 
Pa$\beta$ 1.2822~\micron, and Br$\gamma$ 2.1661~\micron).  In fact, our QSF and 
Kepler spectra resemble the 1.4--2.4~\micron\ spectrum of the optically 
brightest region of RCW~103 presented by OMD90 except that no H$_{2}$ emission 
was seen. 

\begin{figure}[t]
\begin{center}
\includegraphics[scale=0.30,keepaspectratio=true,angle=-90]{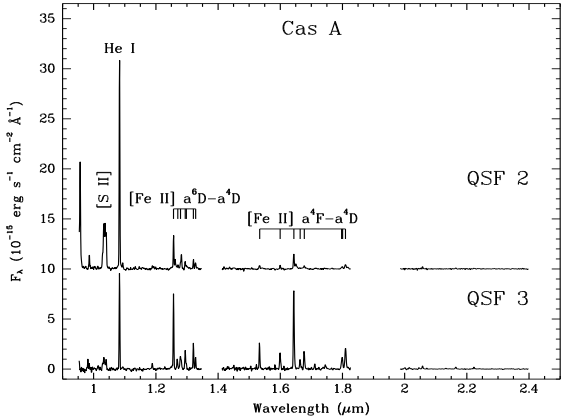}
\caption{Observed NIR spectrum of QSF~2 and QSF~3 in Cas~A,
showing the large variation of the \protect\ion{He}{1}/[\protect\ion{Fe}{2}] emission ratio seen
the the Cas~A QSF data. The QSF~2 data have been shifted vertically by 10 units
for clarity.  \label{QSFcomp}}
\end{center}
\end{figure}

Several of the observed [\ion{Fe}{2}] line ratios can be used as density 
diagnostics (e.g. Nussbaumer \& Storey 1980; OMD89; OMD90).  Using the 
dereddened ratios of the 1.5339~\micron, 1.5999~\micron, and 1.6642~\micron\ 
lines to the strong 1.6440~\micron\ line and the predicted ratios of OMD90, we 
estimated electron densities for each of the three QSFs in Cas~A and for the 
Kepler knot. The results are shown in Table \ref{dentab}.  In each case, the 
various line ratios all gave mutually consistent estimates of the electron 
density for a given knot.  The dereddened 
line ratios in each of the three QSFs observed in Cas~A are indistinguishable to
within the measurement errors and yield electron densities  
of 5 -- 9~$\times$~10$^{4}$~cm$^{-3}$, which approach the 
high-density limit of the [\ion{Fe}{2}] lines.  

The density of the Kepler knot is roughly half that measured in the Cas~A QSFs,
namely 2.5 -- 3.1~$\times$~10$^{4}$~cm$^{-3}$. Our estimate for the density of 
Kepler is somewhat higher than that measured at roughly the same position by 
OMD89, although it might be consistent with their upper limit.  Differences 
in density estimates are not unexpected as OMD89 used a larger 
aperture (5\farcs8~$\times$~5\farcs9 vs.\ our 0\farcs6 slit) and therefore 
sampled a much larger region of the remnant's northwest emission filaments. 

Although electron densities in the three observed Cas~A QSFs are similar, the
\ion{He}{1} to [\ion{Fe}{2}] emission ratio varies dramatically.  This is 
seen in Figure \ref{QSFcomp} which shows the observed NIR spectra
of QSF~2 and QSF~3.  In QSF~2, the \ion{He}{1} 1.083~\micron\ line is much 
brighter than the brightest [\ion{Fe}{2}] lines, while in QSF~3, these lines 
are of nearly equal strength.  Table \ref{dentab} lists the dereddened 
\ion{He}{1} 1.083~\micron\ to [\ion{Fe}{2}] 1.257~\micron\ line ratio which 
varies by nearly a factor of five in the three QSFs observed in Cas~A.   
The observed variation could be due to temperature effects or 
compositional differences in the knots, for instance, in the depletion of 
gaseous iron into dust grains. 

\subsection{\textit{J}- and \textit{K}-band images of Cas~A}

\begin{figure*}[t]
\begin{center}
\includegraphics[scale=0.60,keepaspectratio=true,angle=0]{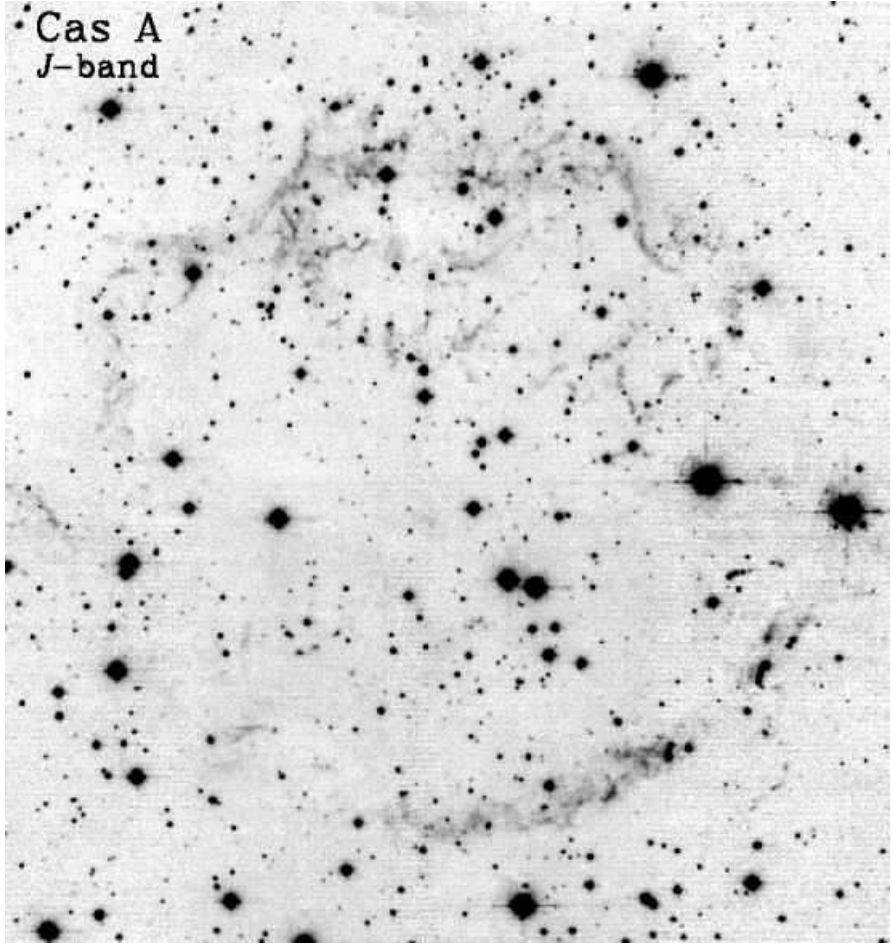}
\caption{\protect\textit{J}-band image of Cas~A.  North is up and
East is to the left.  The \protect\textit{J}-band image is similar to optical
images of Cas~A with both circumstellar knots (QSFs) and ejecta knots (FMKs)
clearly visible.  Diffuse emission is also faintly detected in the
\protect\textit{J}-band near the center of the remnant and around the rim to the north
and west. \label{casajim}}
\end{center}
\end{figure*}

\begin{figure*}[t]
\begin{center}
\includegraphics[scale=0.60,keepaspectratio=true,angle=0]{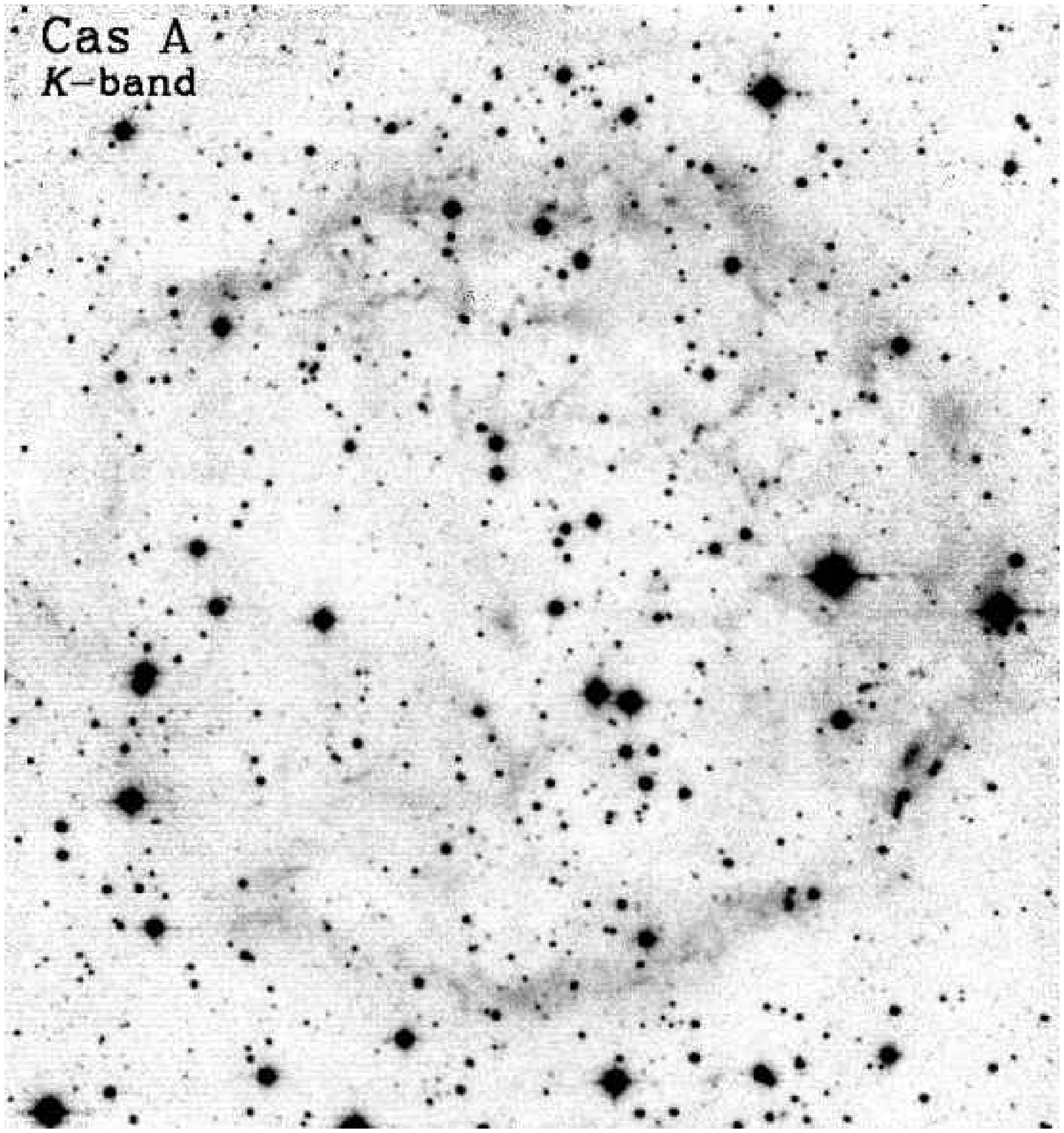}
\caption{\protect\textit{K}-band image of Cas~A.  North is up and
East is to the left.  The dominant feature is diffuse emission that forms a
nearly complete ring.  The diffuse emission exhibits morphological similarities
to radio continuum maps of Cas~A. \label{casakim}}
\end{center}
\end{figure*}

\textit{J}- and \textit{K}-band mosaic images of Cas~A are presented in Figures 
\ref{casajim} and \ref{casakim} respectively.  In both images, the brightest
emission comes from the QSFs.  In \textit{J}-band, this emission is almost
entirely due to strong [\ion{Fe}{2}] lines, while the \textit{K}-band emission
is a mixture of [\ion{Fe}{2}], \ion{He}{1}, and Br$\gamma$.  The relative
brightness of the QSFs in the two bands (i.e. the $J - K$ color) changes from 
knot to knot, and the NIR luminosity is poorly correlated with the optical
luminosity.  This may be related to the large variation of the observed
\ion{He}{1} 1.083~\micron\ to [\ion{Fe}{2}] 1.257~\micron\ line ratio seen in
the NIR spectra of the QSFs.

The FMK ejecta knots are also clearly visible in the \textit{J}- band image.  
In this case, the emission is a mixture of weak [\ion{Fe}{2}] and [\ion{P}{2}]
emission lines.  Some FMK filaments are also visible in the \textit{K}-band
image, particularly in the inner regions of the remnant.  The spectra of the
bright northern FMKs showed no emission in \textit{K}-band, but the 
[\ion{Si}{6}] 1.96~\micron\ line lies at the blue edge of the
\textit{K}-band filter.  Thus the \textit{K}-band filter may be picking
up [\ion{Si}{6}] emission from filaments with large positive radial 
velocities.  This is supported by the fact that the filamentary emission in 
\textit{K}-band largely disappears near the rim of Cas~A, and is seen primarily 
in the inner regions of the remnant where the radial velocities are largest.

The dominant feature in the \textit{K}-band image is diffuse emission which
forms a nearly complete ring around the rim of Cas~A and also fills some of
the interior.  This diffuse emission can be faintly seen in the 
\textit{J}-band image near the center and at several positions on the rim.
This diffuse emission has no optical counterpart and does not correspond well to
X-ray or mid-IR images of Cas~A either (cf. HF96; Lagage et al. 1996), but it 
does exhibit morphological similarities to radio continuum images 
\citep[e.g.][]{ar95}.  Some hint of the morphological differences 
between \textit{K}-band and optical images has been seen in the 2MASS 
images of Cas~A \citep{VD99}.

The lack of any detected diffuse line emission in the
2-D spectra of FMKs and QSFs on the rim (specifically FMKs 1 \& 2, and QSFs
1 \& 2) suggest that the diffuse emission is continuum rather than line 
emission.  The large morphological differences between the \textit{K}-band and 
mid-IR images, and the detection of the diffuse emission in \textit{J}-band 
argue against thermal dust emission as a source.  The thermal dust emission seen
in the mid-infrared is well correlated to the optical knots \citep{L96}, and the
detection of thermal emission in \textit{J}-band would probably require 
prohibitively high dust temperatures at or above typical grain destruction 
limits.  On the other hand, the morphological similarity of the diffuse 
emission with radio continuum images suggests that the diffuse \textit{K}-band
emission may be infrared synchrotron radiation.

\section{Conclusion}
We have presented the first near-infrared (NIR) spectra of a young, metal-rich
supernova remnant.  The spectra of fast-moving ejecta knots in Cas~A
are dominated by [\ion{S}{2}] 1.03~\micron\ emission, but show several other 
faint emission lines including high-ionization lines of [\ion{Si}{6}] and 
[\ion{Si}{10}].  These forbidden silicon lines have often been seen in NIR 
spectra of novae and AGNs but have never before been detected in a supernova 
remnant.  Interestingly, the silicon lines represent a much higher ionization 
state than the observed optical and near-infrared emission lines. Therefore, 
further study of the spatial distribution of the [\ion{Si}{6}] and 
[\ion{Si}{10}] line emission may provide valuable information about the 
ionization structure in the metal-rich Cas~A ejecta.

We also obtained NIR spectra of shocked circumstellar knots in Cas~A and 
Kepler, which are shown to have bright \ion{He}{1} and [\ion{Fe}{2}] emission. 
Analysis of relative [\ion{Fe}{2}] line ratios indicate electron densities 
$\sim$~10$^{4}$ -- 10$^{5}$~cm$^{-3}$ in these knots.  While the measured 
density was relatively constant for the three circumstellar knots observed in 
Cas~A, the \ion{He}{1}/[\ion{Fe}{2}] emission ratios were found to vary by 
nearly a factor of five.  

Finally, we presented  \textit{J} and \textit{K}-band images of Cas~A.
While the \textit{J}-band image is largely similar to optical images, the 
dominant feature in the \textit{K}-band is diffuse emission that best matches
radio continuum images of Cas~A and may be near-infrared synchrotron emission. 
 
\acknowledgments
We would like to thank Bob Barr and the MDM staff for their excellent assistance
in instrument setup and general observing support.  This research is supported 
by NSF Grant 98-76703.

\clearpage
\renewcommand{\arraystretch}{0.8}
\begin{deluxetable}{lcr}
\tablewidth{0em}
\tablecaption{Log of Observations\label{speclog}}
\tablehead{\colhead{Slit} & \colhead{Bandpass} & \colhead{Exposure} \\
	   \colhead{Position}	& \colhead{(\micron)}& \colhead{(s)}}
\startdata
FMK 1 \& 2 	& 0.95 -- 1.8 &	7 $\times$ 900 \\
		& 1.2 -- 2.2  &	3 $\times$ 900 \\
		& 2.0 -- 2.4 &	2 $\times$ 900 \\
FMK 3, 4 \& 5	& 0.95 -- 1.8 & 	6 $\times$ 900 \\
		& 1.2 -- 2.2 &	3 $\times$ 900 \\
QSF 1		& 0.95 -- 1.8 &	5 $\times$ 900 \\
		& 1.2 -- 2.2 &       3 $\times$ 900 \\
		& 2.0 -- 2.4 &	6 $\times$ 900 \\
QSF 2		& 0.95 -- 1.8 &	3 $\times$ 900 \\
		& 1.2 -- 2.2 & 7 $\times$ 900 \\
		& 2.0 -- 2.4 &       3 $\times$ 900 \\
QSF3		& 0.95 -- 1.8 &       3 $\times$ 900 \\
		& 1.2 -- 2.2  &       3 $\times$ 900 \\
		& 2.0 -- 2.4  &	3 $\times$ 900 \\
Kepler		& 0.95 -- 1.8 &       10 $\times$ 900 \\
                & 1.2 -- 2.2  &       5 $\times$ 900 \\
\enddata 
\end{deluxetable}

\begin{deluxetable}{clrrrrrrr}
\tablecolumns{9}
\tablewidth{0em}
\tablecaption{Line Identifications for FMKs 1 \& 2\label{FMK12tab}}
\tablehead{
\colhead{$\lambda _{\rm lab}$} & \colhead{Line} &
\multicolumn{3}{c}{FMK 1} &\colhead{\phd}& \multicolumn{3}{c}{FMK 2} \\
\cline{3-5} \cline{7-9}
\\[-0.6em]
\colhead{(\micron)} & \colhead{ID} & \colhead{$\lambda_{\rm obs}$(\micron)} &
\colhead{I($\lambda$)} & \colhead{F($\lambda$) \tablenotemark{a}} & &
\colhead{$\lambda_{\rm obs}$(\micron)} & \colhead{I($\lambda$)} & 
\colhead{F($\lambda$)\tablenotemark{a}}
}

\startdata
0.9827, 0.9853  & [\ion{C}{1}] $^{3}$P$_{1,2}$--$^{1}$D$_{2}$ & 0.979 & 31 & 
253 & & 0.981 & 58 & 349\\[0.5em]
1.0290, 1.0323, & [\ion{S}{2}] $^{2}$D$_{5/2,3/2}$--$^{2}$P$_{3/2,1/2}$ & 
1.028 & 1280 & 6650 & & 1.027 & 3090 & 11600\\
1.0339, 1.0373\phd \\[0.5em]
1.0824  & [\ion{S}{1}] $^{3}$P$_{2}$--$^{1}$D$_{2}$ & 1.078 & 54 & 250 & & 
1.079 & 228 & 1043 \\
1.0832  & \ion{He}{1} $^{3}$S$_{1}$--$^{3}$P$_{0,1,2}$ \\[0.5em]
1.1289, 1.1290  & \ion{O}{1} $^{3}$P$_{0,1,2}$--$^{3}$D$_{1,2,3}$ & 1.125 & 33 
& 139 & & 1.125 & 55 & 238 \\
1.1309  & [\ion{S}{1}] $^{3}$P$_{1}$--$^{1}$D$_{2}$ \\[0.5em]
1.1471  & [\ion{P}{2}] $^{3}$P$_{1}$--$^{1}$D$_{2}$ & 1.143 & 20 & 79 & & 
1.142 & 39 & 160 \\
1.1886  & [\ion{P}{2}] $^{3}$P$_{2}$--$^{1}$D$_{2}$ & 1.184 & 50 & 186 & & 
1.183 & 103 & 375 \\
1.2570  & [\ion{Fe}{2}] a$^{6}$D$_{9/2}$--a$^{4}$D$_{7/2}$ & 1.251 & 37 & 120 
& & 1.254 & 76 & 253 \\
1.2791  & [\ion{Fe}{2}] a$^{6}$D$_{3/2}$--a$^{4}$D$_{3/2}$ &\nodata&\nodata&
\nodata& & 1.279 & 21 & 67 \\
1.2946  & [\ion{Fe}{2}] a$^{6}$D$_{5/2}$--a$^{4}$D$_{5/2}$ &\nodata&\nodata&
\nodata& & 1.293 & 23 & 66 \\
1.3209  & [\ion{Fe}{2}] a$^{6}$D$_{7/2}$--a$^{4}$D$_{7/2}$ &\nodata&\nodata&
\nodata& & 1.319 & 17 & 50 \\
1.4305  & [\ion{Si}{10}] $^{2}$P$_{1/2}$--$^{2}$P$_{3/2}$ & 1.425 & 35 & 93 & &
1.424 & 76 & 111 \\[0.5em]
1.6440  & [\ion{Fe}{2}] a$^{4}$F$_{9/2}$--a$^{4}$D$_{7/2}$ & 1.639 & 27 & 59& &
1.638 & 54 & 118 \\
1.6459  & [\ion{Si}{1}] $^{3}$P$_{2}$--$^{1}$D$_{2}$ \\[0.5em]
1.965   & [\ion{Si}{6}] $^{2}$P$_{3/2}$--$^{2}$P$_{2/2}$ & 1.955 & 69 & 124 & &
1.952 & 268 & 480 
\enddata
\tablecomments{Line fluxes in units of \protect$10^{-15}$ erg s\protect$^{-1}$ cm\protect$^{-2}$}
\tablenotetext{a}{Corrected for \protect\textit{E(B--V)} = 1.5}
\end{deluxetable}

\begin{deluxetable}{clrrrrrrrrrrr}
\tablecolumns{13}
\tablewidth{57em}
\rotate
\tablecaption{Line Identifications for FMKs 3, 4, \& 5\label{FMK345tab}}
\tablehead{
\colhead{$\lambda _{\rm lab}$} & \colhead{Line} & \multicolumn{3}{c}{FMK 3}&
\colhead{\phd}&
\multicolumn{3}{c}{FMK 4}&\colhead{\phd}&\multicolumn{3}{c}{FMK 5} \\
\cline{3-5} \cline{7-9} \cline{11-13}
\\[-0.6em]
\colhead{(\micron)} & \colhead{ID} & \colhead{$\lambda_{\rm obs}$(\micron)} &
\colhead{I($\lambda$)} & \colhead{F($\lambda$)\tablenotemark{a}} & & 
\colhead{$\lambda_{\rm obs}$(\micron)} &
\colhead{I($\lambda$)} & \colhead{F($\lambda$)\tablenotemark{a}} & & 
\colhead{$\lambda_{\rm obs}$(\micron)} &
\colhead{I($\lambda$)} & \colhead{F($\lambda$)\tablenotemark{a}}
}
\startdata
0.9827, 0.9853  & [\ion{C}{1}] $^{3}$P$_{1,2}$--$^{1}$D$_{2}$ & 0.977  & 22  & 
128 & & 0.978 & 51 & 306 & & \nodata & \nodata & \nodata\\[0.5em]
1.0290, 1.0323, & [\ion{S}{2}] $^{2}$D$_{5/2,3/2}$--$^{2}$P$_{3/2,1/2}$ & 
1.025 & 464 & 2450 & & 1.025 & 412 & 2210 & & 1.044 & 156 & 784 \\
1.0339, 1.0373\phd \\[0.5em]
1.0824  & [\ion{S}{1}] $^{3}$P$_{2}$--$^{1}$D$_{2}$ & 1.075 & 41  & 186 & &
1.075 & 30 & 139 & & 1.096 & 15 & 61 \\
1.0832  & \ion{He}{1} $^{3}$S$_{1}$--$^{3}$P$_{0,1,2}$ \\[0.5em]
1.1289, 1.1290  & \ion{O}{1} $^{3}$P$_{0,1,2}$--$^{3}$D$_{1,2,3}$ & 1.122 & 31 
& 135 & & 1.122 & 25 & 106 & & 1.144 & 6 & 24 \\
1.1309  & [\ion{S}{1}] $^{3}$P$_{1}$--$^{1}$D$_{2}$ \\[0.5em]
1.1886  & [\ion{P}{2}] $^{3}$P$_{2}$--$^{1}$D$_{2}$ & 1.180 & 19 & 70 & &1.181
& 23 & 87 & & 1.202 & 11 & 39 \\
1.2570  & [\ion{Fe}{2}] a$^{6}$D$_{9/2}$--a$^{4}$D$_{7/2}$ & 1.248 & 16 & 57& &
1.248 & 16 & 60 & &1.271 & 20 & 64 \\
1.4305  & [\ion{Si}{10}] $^{2}$P$_{1/2}$--$^{2}$P$_{3/2}$ &\nodata&\nodata&
\nodata& &1.420 & 16 & 40 & &\nodata & \nodata & \nodata \\[0.5em]
1.6440  & [\ion{Fe}{2}] a$^{4}$F$_{9/2}$--a$^{4}$D$_{7/2}$ & 1.632 & 14  & 30 
& & 1.633 & 20 & 43 & & 1.663 & 20 & 44\\
1.6459  & [\ion{Si}{1}] $^{3}$P$_{2}$--$^{1}$D$_{2}$ \\[0.5em]
1.965   & [\ion{Si}{6}] $^{2}$P$_{3/2}$--$^{2}$P$_{2/2}$ &\nodata&\nodata&
\nodata& &1.951 & 20 & 36 & &\nodata & \nodata & \nodata
\enddata
\tablecomments{Line fluxes in units of \protect$10^{-15}$ erg s\protect$^{-1}$ cm\protect$^{-2}$}
\tablenotetext{a}{Corrected for \protect\textit{E(B--V)} = 1.5}
\end{deluxetable}

\begin{deluxetable}{clrrrrrrrrrrr}
\tablecolumns{13}
\rotate
\tablewidth{52em}
\tablecaption{Line Identifications for Cas A QSFs and Kepler SNR\label{QSFkeptab}}
\tablehead{
~\\[-0.8em]
\colhead{$\lambda _{\rm rest}$} & \colhead{Line} & 
\multicolumn{2}{c}{Cas A QSF 1} & \colhead{\phd} & 
\multicolumn{2}{c}{Cas A QSF 2} & \colhead{\phd} & 
\multicolumn{2}{c}{Cas A QSF 3} & \colhead{\phd} & 
\multicolumn{2}{c}{Kepler} \\
\cline{3-4} \cline{6-7} \cline{9-10} \cline{12-13}
\\[-0.8em]
\colhead{(\micron)} & \colhead{ID} & 
\colhead{I($\lambda$)} & 
\colhead{F($\lambda$)\tablenotemark{a}} & \colhead{} &
\colhead{I($\lambda$)} & 
\colhead{F($\lambda$)\tablenotemark{a}} & \colhead{} &
\colhead{I($\lambda$)} & 
\colhead{F($\lambda$)\tablenotemark{a}} & \colhead{} &
\colhead{I($\lambda$)} & 
\colhead{F($\lambda$)\tablenotemark{b}}
}
\startdata
0.9827, 0.9853  & [\ion{C}{1}] $^{3}$P$_{1,2}$--$^{1}$D$_{2}$ &\nodata&\nodata& 
& 44 & 253 & & 36 & 220 & & 49 & 145 \\[0.5em]
1.0290, 1.0323, & [\ion{S}{2}] $^{2}$D$_{5/2,3/2}$--$^{2}$P$_{3/2,1/2}$ & 111 & 
595 & & \multicolumn{2}{c}{(bl w/ FMK)}  & & 104 & 534 & & 157 & 423 \\
1.0339, 1.0373\phd \\[0.5em]
1.0832  & \ion{He}{1} $^{3}$S$_{1}$--$^{3}$P$_{0,1,2}$ & 450 & 2060 & & 607 & 
2780 & & 276 & 1260 & & 340 & 848 \\
1.0941  & \ion{H}{1} Pa$\gamma$ & 16 & 71 & & 14 & 65 & & 9 & 37 & & 17 & 42 \\
1.1885  & [\ion{Fe}{2}] a$^{4}$D$_{7/2}$--a$^{2}$G$_{7/2}$ & 12 & 45 & &\nodata&
\nodata & & 20 & 72 & & 18 & 42 \\
1.2570  & [\ion{Fe}{2}] a$^{6}$D$_{9/2}$--a$^{4}$D$_{7/2}$ & 173 & 572 & & 95 & 
315 & & 211 & 700 & & 329 & 675 \\
1.2707  & [\ion{Fe}{2}] a$^{6}$D$_{1/2}$--a$^{4}$D$_{1/2}$ & 27 & 83 & & 17 & 61& & 37 & 120 & & 31 & 63\\[0.5em]
1.2791  & [\ion{Fe}{2}] a$^{6}$D$_{3/2}$--a$^{4}$D$_{3/2}$ & 85 & 269 & & 65 & 
210 & & 78 & 250 & & 79 & 160 \\
1.2822  & \ion{H}{1} Pa$\beta$ \\[0.5em]
1.2946  & [\ion{Fe}{2}] a$^{6}$D$_{5/2}$--a$^{4}$D$_{5/2}$ & 60 & 185 & & 39 & 
118 & & 70 & 218 & & 82 & 162\\
1.2981  & [\ion{Fe}{2}] a$^{6}$D$_{1/2}$--a$^{4}$D$_{3/2}$\\[0.5em]
1.3209  & [\ion{Fe}{2}] a$^{6}$D$_{7/2}$--a$^{4}$D$_{7/2}$ & 58 & 176 & & 28 & 
83 & & 75 & 226 & & 84 & 172 \\
1.3281  & [\ion{Fe}{2}] a$^{6}$D$_{3/2}$--a$^{4}$D$_{5/2}$ & 32 & 97 & & 28 & 
87 & & 36 & 107 & & 35 & 72 \\
1.5339  & [\ion{Fe}{2}] a$^{4}$F$_{9/2}$--a$^{4}$D$_{5/2}$ & 58 & 138 & & 17 & 
38 & & 79 & 187 & & 61 & 105 \\
1.5999  & [\ion{Fe}{2}] a$^{4}$F$_{7/2}$--a$^{4}$D$_{3/2}$ & 46 & 104 & & 12 & 
27 & & 67 & 151 & & 44 & 71 \\
1.6440  & [\ion{Fe}{2}] a$^{4}$F$_{9/2}$--a$^{4}$D$_{7/2}$ & 203 & 440 & & 57 & 
125 & & 288 & 628 & & 273 & 432 \\
1.6642  & [\ion{Fe}{2}] a$^{4}$F$_{5/2}$--a$^{4}$D$_{1/2}$ & 28 & 61 & &\nodata&
\nodata & & 39 & 80 & & 29 & 45 \\
1.6773  & [\ion{Fe}{2}] a$^{4}$F$_{7/2}$--a$^{4}$D$_{5/2}$ & 44 & 93 & & 18 & 
44 & & 63 & 134 & & 45 & 69 \\[0.5em]
1.7976  & [\ion{Fe}{2}] a$^{4}$F$_{3/2}$--a$^{4}$D$_{3/2}$ & 54 & 108 & & 14 & 
27 & & 68 & 134 & & 70 & 99 \\
1.8005  & [\ion{Fe}{2}] a$^{4}$F$_{5/2}$--a$^{4}$D$_{5/2}$\\[0.5em]
1.8099  & [\ion{Fe}{2}] a$^{4}$F$_{7/2}$--a$^{4}$D$_{7/2}$ & 59 & 109 & & 30 & 
53 & & 109 & 211 & & 57 & 91\\
2.0466  & [\ion{Fe}{2}] a$^{4}$P$_{5/2}$--a$^{2}$P$_{3/2}$ & 9 & 16 & &\nodata&
\nodata & & 7 & 12 & &\nodata & \nodata\\
2.0587  & \ion{He}{1} $^{1}$S$_{0}$--$^{3}$P$_{1}$ & 20 & 35 & & 8 & 13 & & 13 &
22 & & \nodata & \nodata\\
2.1661  & \ion{H}{1} Br$\gamma$ & 12 & 20 & & 5 & 8 & & 9 & 16 & &\nodata & 
\nodata \\
2.2244  & [\ion{Fe}{2}] a$^{4}$G$_{9/2}$--a$^{2}$H$_{11/2}$ & 15 & 23 & & 5 & 8 
& & 8 & 12 & & \nodata & \nodata
\enddata
\\[-2em]
\tablecomments{Line fluxes in units of \protect$10^{-15}$ erg s\protect$^{-1}$ cm\protect$^{-2}$.}
\tablenotetext{a}{Corrected for \protect\textit{E(B--V)} = 1.5}
\tablenotetext{b}{Corrected for \protect\textit{E(B--V)} = 0.9}
\end{deluxetable}

\begin{deluxetable}{lccccc}
\tablewidth{0em}
\tablecolumns{6}
\tablecaption{Cas~A QSF \& Kepler: Knot Densities \& \protect\ion{He}{1}/[\protect\ion{Fe}{2}] 
Flux Ratios\label{dentab}}
\tablehead{
\colhead{Knot}& 
\colhead{\underline{[\ion{Fe}{2}] $\lambda$1.5339}}&  
\colhead{\underline{[\ion{Fe}{2}] $\lambda$1.5999}}& 
\colhead{\underline{[\ion{Fe}{2}] $\lambda$1.6642}}&
\colhead{$n_{e}$}& 
\colhead{\underline{[\ion{He}{1}] $\lambda$1.0832}}\\
\colhead{} & \colhead{[\ion{Fe}{2}] $\lambda$1.6440} &
\colhead{[\ion{Fe}{2}] $\lambda$1.6440} & 
\colhead{[\ion{Fe}{2}] $\lambda$1.6440} & 
\colhead{($10^{4}$ cm$^{-3}$)} & 
\colhead{[\ion{Fe}{2}] $\lambda$1.2570}
\\[-0.7em]
}
\startdata
Cas A QSF 1 & $0.314 \pm 0.015$ & $0.236 \pm 0.015$ & $0.139 \pm 0.015$ &
6.0 -- 8.7 & $3.6 \pm 0.2$ \\
Cas A QSF 2 & $0.304 \pm 0.040$ & $0.216 \pm 0.040$ & \nodata &
3.2 -- 8.9 & $8.8 \pm 0.4$\\
Cas A QSF 3 & $0.298 \pm 0.010$ & $0.240 \pm 0.010$ & $0.127 \pm 0.010$ &
5.2 -- 6.9 & $1.8 \pm 0.1$\\
Kepler SNR  & $0.243 \pm 0.015$ & $0.164 \pm 0.015$ & $0.104 \pm 0.015$ &
2.5 -- 3.1 & $1.3 \pm 0.1$\\
\enddata
\end{deluxetable}

\end{document}